\documentclass{PoS}

\title{ Dark Matter in galaxies: leads to its Nature}

\ShortTitle{DM  in galaxies}

\author{\speaker{Paolo Salucci}\thanks{}\\
     SISSA, Via Bonomea  265, Trieste, Italy\\
        E-mail: \email{salucci@sissa.it}}

\author{Mariafelicia De Laurentis\\
       Dipartimento di Scienze Fisiche, Universit�
di Napoli {}``Federico II'' and INFN sez. di Napoli, Compl. Univ. di
Monte S. Angelo, Edificio G, Via Cinthia, I-80126, Napoli, Italy\\
        E-mail: \email{felicia@na.infn.it}}

\abstract{ Recent observations have revealed the structural properties of the dark and luminous mass distribution in spirals.
 These results led to the vision of a new and amazing scenario.
The investigation of single and coadded objects has shown that the rotation curves of  spirals  follow, from their centers out to  their virial radii, an universal profile that implies a tuned  combination of  their stellar disk and dark halo mass distributions. This,  alongside with accurate mass modeling of individual galaxies, poses important challenges to the   presently  theoretically favored $\Lambda$CDM Cosmology.   }

\FullConference{VIII International Workshop on the Dark Side of the Universe,\\
		June 10-15, 2012\\
		Rio de Janeiro, Brazil}

\begin{document}

\section{Introduction}
The presence of wide content of invisible matter in and around spiral galaxies, distributed differently from stars and gas, is well fixed from optical and $21$ cm rotation curves (RCs) which do not show the expected Keplerian fall-off at large radii but remain increasing, flat or only slightly decreasing over their entire observed range \cite{rubin80,Bosma81}.  The extra  mass component becomes progressively more abundant at outer radii and in  the less luminous galaxies \cite{Persic}. 

The circular velocity $V(r) $ of  spiral galaxies,  is the equilibrium velocity due to their mass distribution. The gravitational potentials of the spiral's mass components and namely those of  a spherical stellar bulge, a dark matter (DM)  halo, a stellar disk and a gaseous disc  
$\phi_{tot}=\phi_b+ \phi_{DM}+\phi_*+\phi_{HI}$
lead to:
\begin{equation}
V^2_{tot}(r)=r\frac{d}{dr}\phi_{tot}=V^2_b + V^2_{DM}+V^2_*+V^2_{HI}. \label{eq:3_vel_tot}
\end{equation}
with  the Poisson equation  relating the surface/spatial  densities to the corresponding  gravitational potentials. 
 The surface stellar  density  $\Sigma_*(r)$, 
is assumed proportional to the luminosity 
surface density  \cite{Freeman1970}, so that:
\begin{equation}
\Sigma_{*}(r)=\frac{M_{D}}{2 \pi R_{D}^{2}}\: e^{-r/R_{D}},\label{eq:3_sigma_stars}
\end{equation}
where $M_D$ is the disk mass and $R_D$ is the scale length,  $\Sigma_{HI}(r)$  is   directly derived  by HI flux observations. From the above it follows:  

\begin{equation}V_{*}^{2}(r)=\frac{G M_{D}}{2R_{D}} x^{2}B\left(\frac{x}{2}\right)\end{equation}
where $x\equiv r/R_{D}$, $G$ is the gravitational constant   $B=I_{0}K_{0}-I_{1}K_{1}$,  a combination of Bessel functions \cite{Freeman1970}.

\begin{figure} [h!]
\centering
\vskip -1.1cm
\includegraphics[width=15cm]{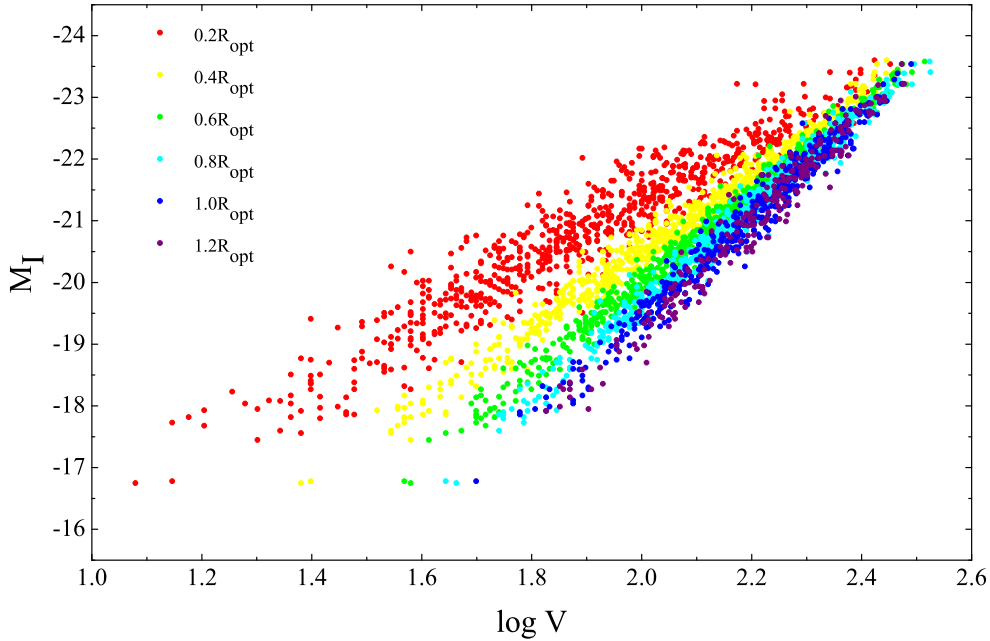}
\vskip -1.0cm
\caption{The Radial Tully-Fisher. The relations at different radii are indicated with different colours.} 
\label{fig:RTF}
\end{figure}

The  rotation curve  of a  spiral  is   a fair  measure of its gravitational potential.
In fact: \begin{itemize}
\item in their  very inner regions  the light  well  traces the gravitating  mass \cite{Ratnam} and 
\item  there exists,  at any  galactocentric radii measured in units of disk length-scale $R_n \equiv (n/5)R_{opt}$,  a    {\it radial}  Tully-Fisher relation \cite{Yegorova} linking,  with  very low scatter,  the local rotation  velocity  $V_n \equiv V_{rot}(R_n)$ and the  total galaxy magnitude $M$   (see Fig. \ref{fig:RTF}).
\begin{equation}
M = a_n \log V_n + b_n , \label{eq:RTF}
\end{equation} 
($a_n$, $b_n$ are the slope and zero-point of the relations)
 .\end{itemize}

In Fig. 2 of   \cite{PSS} it is shown, for a  large  sample of galaxies, $\nabla$, the  logarithmic   slope  of the circular velocity   at $R_{opt} $ as a function  of $V_{opt}$ and $M_B$. One finds:  $ -0.2 \leq \nabla \leq 1$: at this radius, the  RCs slopes  take almost  all  the values allowed by Newtonian gravity,   from -0.5 (Keplerian regime) to  1 (solid body regime);  furthermore,   $\nabla$ strongly correlates  with   galaxy  luminosity and   $V_{opt}$  \cite{PSS}. On the other hand,  spirals show  an inner baryon dominance region  whose size ranges   between  1 and 3 disk exponential length-scales according to the galaxy
luminosity (see Fig(8) of \cite{PSS} and  \cite{Athanassoula,Palunas}). Inside this region, 
  the  ordinary baryonic matter fully accounts for the  rotation curve;   outside it, instead,    
it cannot justify the observed {\it profile} and   {\it amplitude}.

\begin{figure} [h!] 
\centering
%%\vskip 1.6cm
%%\hskip -6.9cm
\includegraphics[width=12cm]{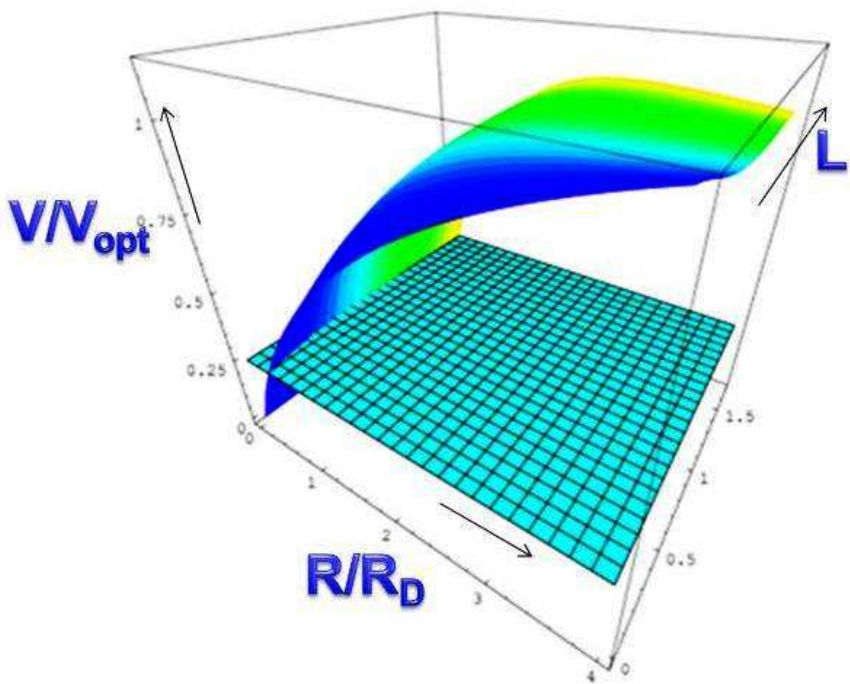}
\vskip -0.5truecm
\caption{The Universal rotation curve of spiral galaxies.} 
\label{fig:new}
\vskip 0.3truecm
\end{figure}

This riddle  is  solved by adding an extra mass component, a dark matter (DM) halo, whose existence and properties are assessed by the RC's and  whose  phenomenology  is  evident in  the 11 coadded rotation curves $V_{coadd}(r/R_{opt}, M_I)$  obtained  by  binning  $~1000$ RCs of late type spirals extended  out to $R_{opt} \equiv 3 R_D $. These synthetic  RCs  correspond to spirals  of   luminosities  spanning their whole $I$-band range: $-16.3 < {\rm M}_I< -23.4$.   This  led to  the  Universal Rotation Curve (URC) paradigm:  there exists  a function of radius and luminosity that well fit the RC of any spiral galaxy~(see \cite{Salucci07}  and references therein). Additional  kinematics  data,  including    very extended  individual RCs  and   virial velocities $V_{vir}\equiv (G M_{vir}/R_{vir})^{1/2}$ obtained in  \cite{shankar06},  further  support the  URC paradigm and  accurately determine the Universal velocity function  out to the virial radius\cite{Salucci07}. Then, $V^2_{URC} = V^2_{URCD} + V^2_{URCH}$  becomes the  observational counterpart of the  rotation curves  of spirals  emerging  out of  cosmological simulations  (e.g. \cite{nfw96}).

\begin{figure}
\begin{center}
\vskip -5.9truecm
\includegraphics[width=15truecm]{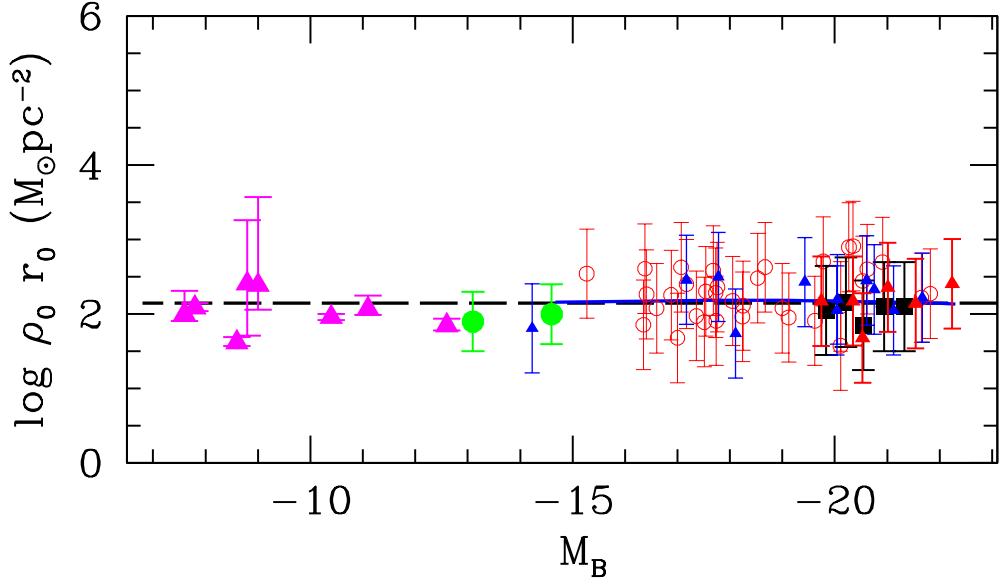}
\end{center}
\vskip -1.7truecm
\caption{$\rho_0 r_0$  as a function of  galaxy magnitude and  Hubble type. Data:
  \cite{spano08} sample of spiral galaxy data (open
  red circles); URC relation (solid blue
  line) \cite{shankar06};  dwarf irregulars N 3741~  \cite{gentile2007} and DDO
  $47~ $; \cite{gentile05} (green full circles);  ellipticals investigated by weak lensing  \cite{donato09} (black squares); Milky Way dSphs  (pink triangles); spirals in THINGS sample~ (blue triangles); \cite{Walter08};  early-type
  spirals~ (red triangles); \cite{Noor07}.  Long-dashed line shows the \cite{donato09} result.}
\label{fig:sigmacost}
\end{figure}

To model the URC (and any individual RC) we assume the  Burkert profile for the  DM halo \cite{sb00} :

\begin{equation}
\rho (r)={\rho_0\, r_0^3 \over (r+r_0)\,(r^2+r_0^2)}~,\label{BH}
\end{equation}
$\rho_0$ and $r_0$ and    $M_D$ (see above)  are, respectively, the  DM central density, its core radius  
and the galaxy disk mass: these 3 free structural parameters  get determined  by $\chi^2$ fitting the  rotation curves.  Remarkably, for all available  kinematics of thousands spirals, the above mass  model fits  data in  excellent way \cite{Salucci07}. 

At any radii, objects  with lower luminosity  have larger  dark-to-stellar mass ratio and   denser DM halos, with their  central values spanning two order of magnitudes over the  whole mass sequence of spirals.

\begin{figure}[!h]
\centering
\vskip -0.3cm
\includegraphics[width=13truecm]{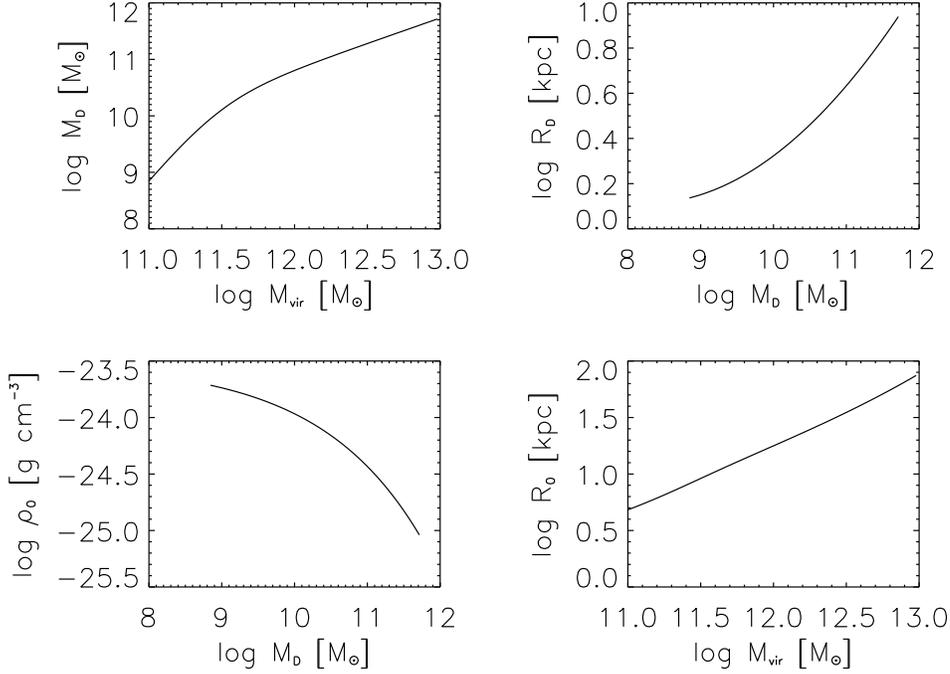}
\vskip -0.7truecm 
\caption{Scaling relations between the structural parameters of spirals.}
\label{fig:scaling_relations}
\vskip 0.3cm
\end{figure}

Furthermore,  a number   of  scaling laws among  the structural mass parameters $\rho_0$, $M_D$, $M_{vir}$, $r_0$ emerges  (see  Fig.\ref{fig:scaling_relations} taken from \cite{Salucci07}). Among these, we must draw attention on  the quantity $\mu_{0D}\equiv \rho_0 r_0$,
proportional to the halo central surface density, that it has  been  found  independent of the galaxy  magnitude and Hubble Type   
  
\begin{equation}
{\rm log} \frac{\mu_{0D}}{\rm M_{\odot} pc^{-2}} = 2.2 \pm 0.25                                                  
 \end{equation}
This relationship pioneered in    \cite{Kormendy},  is  further supported by   \cite{spano08} and  \cite{donato04}.

The  relationships between global galaxy quantities are also  important. First, the halo mass is the fundamental physical quantity characterizing a spiral galaxy. Its halo mass lies in the  range $3 \times 10^{10} M_{\odot} \leq  M_h\leq  3\times10^{13} M_{\odot} $.  Halos of very   low mass $< \times 10^{10} M_{\odot}  $,  hosting disk systems,   are not detected.  Halos of very large mass $ >3\times10^{13} M_{\odot} $ host groups of galaxies rather than a single object.  

\cite{Shankar} by  means of abundance matching method derived  relations between virial halo masses ($M_h$) and galaxy properties, including $r*$-band luminosities ($L_r$) and  stellar component masses ($M_{star}$) .

\begin{equation}\label{eq|MhLr}
%{M_h}\approx 3\times 10^{11}\, M_{\odot}\, \left(\frac
%{L_{r}}{1.3\times 10^{10}\, L_{\odot}}\right)^{0.35}
%\left[1+\left(\frac {L_{r}}{1.3\times 10^{10}\,
%L_{\odot}}\right)^{1.30}\right]~.
{M_h\over 3\times 10^{11}\, M_{\odot}} = \left[\left(\frac
{L_{r}}{1.3\times 10^{10}\, L_{\odot}}\right)^{0.35}+\left(\frac
{L_{r}}{1.3\times 10^{10}\, L_{\odot}}\right)^{1.65}\right]~.
\end{equation}

These relationships, i.e.  eq (1.7) and  eqs. (12)-(14) of  \cite{Shankar}), are well represented by double power laws, with a break at $M_{h,break} \approx 3\times 10^{11} M_{\odot}$, corresponding to a mass in stars $M_{star}\sim  1.2\times10^{10} M_{\odot}$ and to an $r*$-band luminosity $L_r\sim  5\times10^9 L_{\odot}$.

In \cite{evoli} the relationship in late-type galaxies between the neutral hydrogen (HI) disk mass and the stellar disk mass     has been  derived by abundance matching the stellar disk mass function from the Sloan Digital Sky Survey and the HI mass function from the HI Parkes All Sky Survey (HIPASS). As result,  the HI mass in late-type galaxies tightly correlates with the stellar mass over three orders of magnitude in stellar disk mass (see eq.(5) in \cite{evoli}).

 Remarkably, the baryonic fraction in a spiral  is much smaller than the cosmological value $\Omega_b/\Omega_{matter} \simeq 1/6  $, and it ranges between $7\times 10^{-3}$   to   $5\times 10^{-2}$, (see Fig. 3 in \cite{evoli}). On the other hand the mass transformed in stars is a strong function of the  halo mass  suggesting that processes such as  Supernovae (SN)  explosions  must have heated up a very large fraction of the original hydrogen.

The above discussed relations bear the imprint of the processes ruling galaxy formation, and highlight the inefficiency of galaxies both in forming stars below a typical  mass $M_{h,break}$   marking the maximum efficiency of the star forming process.

In spirals there is  fundamental evidence that  dark and  luminous matter  are well linked together, would  this be  the imprint of the nature of the DM itself?

\section{The DM core-cusp issue}

The cuspiness of the DM halos density profiles plays a  central role in Cosmology. In fact, a cuspy density profile   is predicted   by  (the simplest version of)  the  currently favored  Cold Dark Matter (CDM) scenario, in detail, from the outcome    of high-resolution numerical  simulations of the  structure formation \cite{nfw96,moore1999}. Surprisingly, however, such a cusp it is not seen
in real kinematical data (e.g. \cite{gentile05,deblok1, Marchesini02,gentile04,gentile2007}), that, in addition, show   unexplained systematics in  the DM distribution (see \cite{donato09}).

Let us recall that in $\Lambda CDM $ the halo  spatial density  is found  universal and   well reproduced by one-parameter  radial profile \cite{nfw96}:

\begin{equation}
\rho_{NFW}(r) = \frac{\rho_s}{(r/r_s)\left(1+r/r_s\right)^2},
\label{eq:nfw}
\end{equation}

where $r_s$ is a characteristic inner radius, and $\rho_s$ the corresponding density.  The virial radius  $R_{vir}$ and  halo mass $M_{vir}$  the mean  universal  density $\rho_u$ are related by: $
M_{vir}  \simeq   100  \rho_u R_{vir}^3.$ Numerical simulations show also that $r_s$ and $\rho_s$  are related within a reasonable scatter:    $ R_{vir} / r_s \simeq 9.7 \left( \frac{M_{vir}}{10^{12}M_{\odot}} \right)^{-0.13}$. 

Since  they were found in numerical simulations, the cuspy NFW density profiles   disagreed with the actual profiles of dark matter  halo    around   spirals and LSB \cite{moore94,kravtsov98,McGaugh98}. However, strong concerns were raised  that this early evidence  was  biased by observational  systematics. The claim  that the observed apparent cores rather  signaled  an hidden cusp was  frequently put forward. The solution of  this riddle, that  lies in   the very nature of Dark Matter,  was found by properly   investigating a number of  suitable test-cases. That is,  by careful modeling   2D,   high quality,  extended, regular, free from deviations from axial symmetry  rotation curves that were trustable  up to their second spatial derivatives. \cite{salucci03,gentile04}. In addition, a step forward has come from  adopting,  for the Dark matter halo,   the  Burkert profile (see eq. (\ref{BH}));  this profile is cored at small radii, but it  converges  to   the NFW profile  for  $r >0.3  R_{vir}$. As result of this, the RCs data themselves,   by determining in an unbiased way the value of $r_0$, are able to  define the actual level of the DM halo cuspiness in Spirals.

About a decade of investigations can be summarized as it follows:   in  all examined  cases,  NFW halo predictions and observations  are  in plain disagreement on {\it several}  aspects: the disk + NFW halo mass model 
\begin{itemize}
\item fits the RCs  poorly and
\item  implies  an implausibly  low stellar mass-to-light ratio  and in some case
\item an unphysical  high halo mass 
\end{itemize}

(see e.g. \cite{spekkens05, gentile04, gentile05, Gentile07}).
\begin{figure}[h!] 
\centering
\vskip -0.7cm
\hskip -7.5cm
\includegraphics[width=9.5cm]{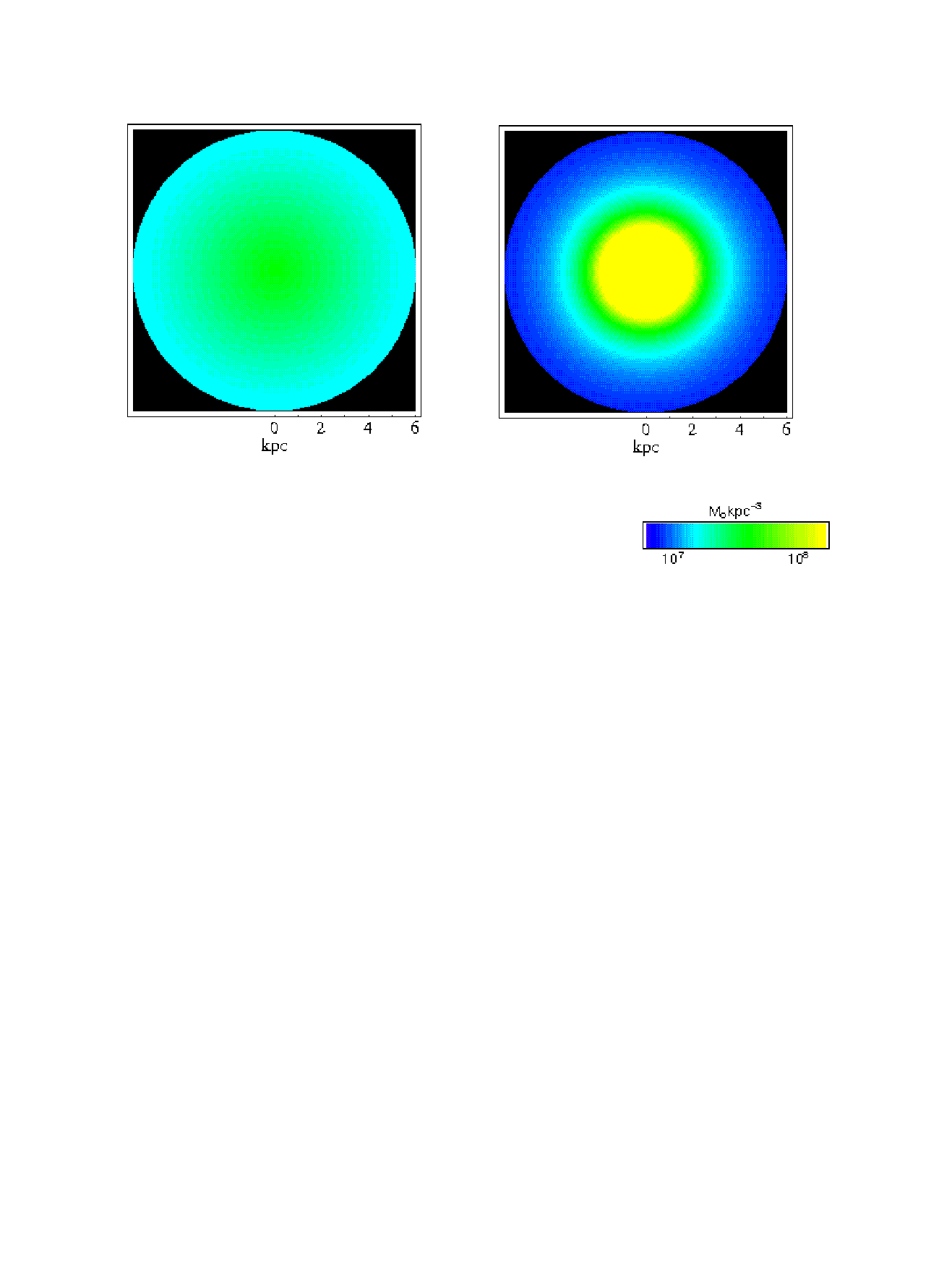}
\vskip -13.0cm
\hskip 7cm
\includegraphics[width=6.2cm]{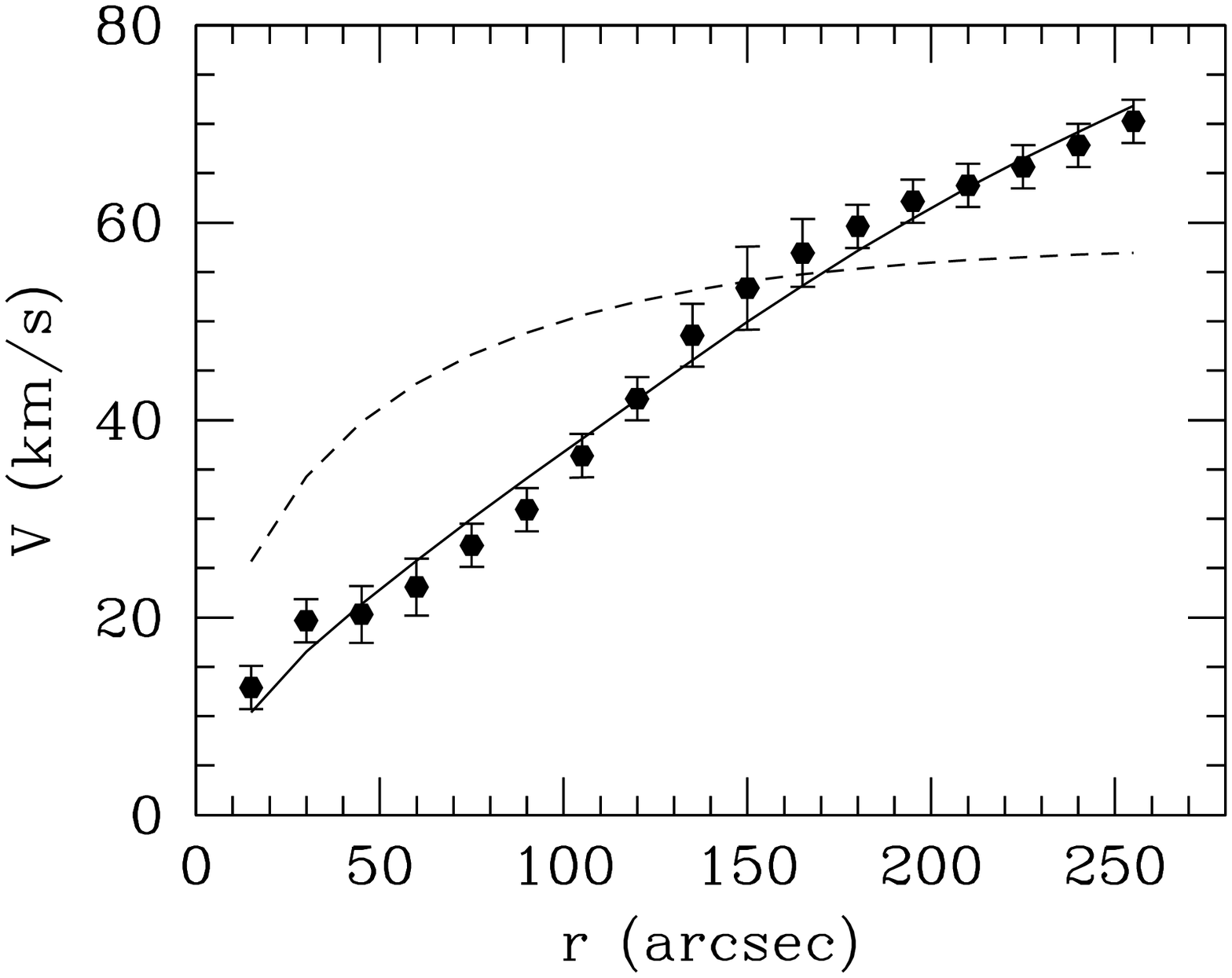}
\vskip 0.5cm
\caption{Left: Dark halo density in ESO 116-G12: observations (left)  vs. CDM predictions (right) \cite{gentile04}. Right: DDO 47 RC best-fits: URC model (solid line) compared with  NFW  halo + stellar disk (dashed line)  \cite{gentile05}}
\label{fig:cuspVScore}
\end{figure} 

It is worth  illustrating a couple of example of this disagreement: the galaxy ESO $116$-G$12$,  see Fig 5 (left) and   the nearby spiral dwarf galaxy DDO 47 see Fig 5 (right). For the latter, the RC mass modeling finds that the dark halo density has a core about $7$ kpc wide and a central density $\rho_0 = 1.4 \times 10^{-24}$ g cm$^{-3}$:  this density profile is  {\it much} shallower  than that predicted by a NFW profile that results totally unable to fit the RC. 

\begin{figure}[!h] 
\vskip 0.2cm
\hskip 0.cm
\includegraphics[width=11.4cm]{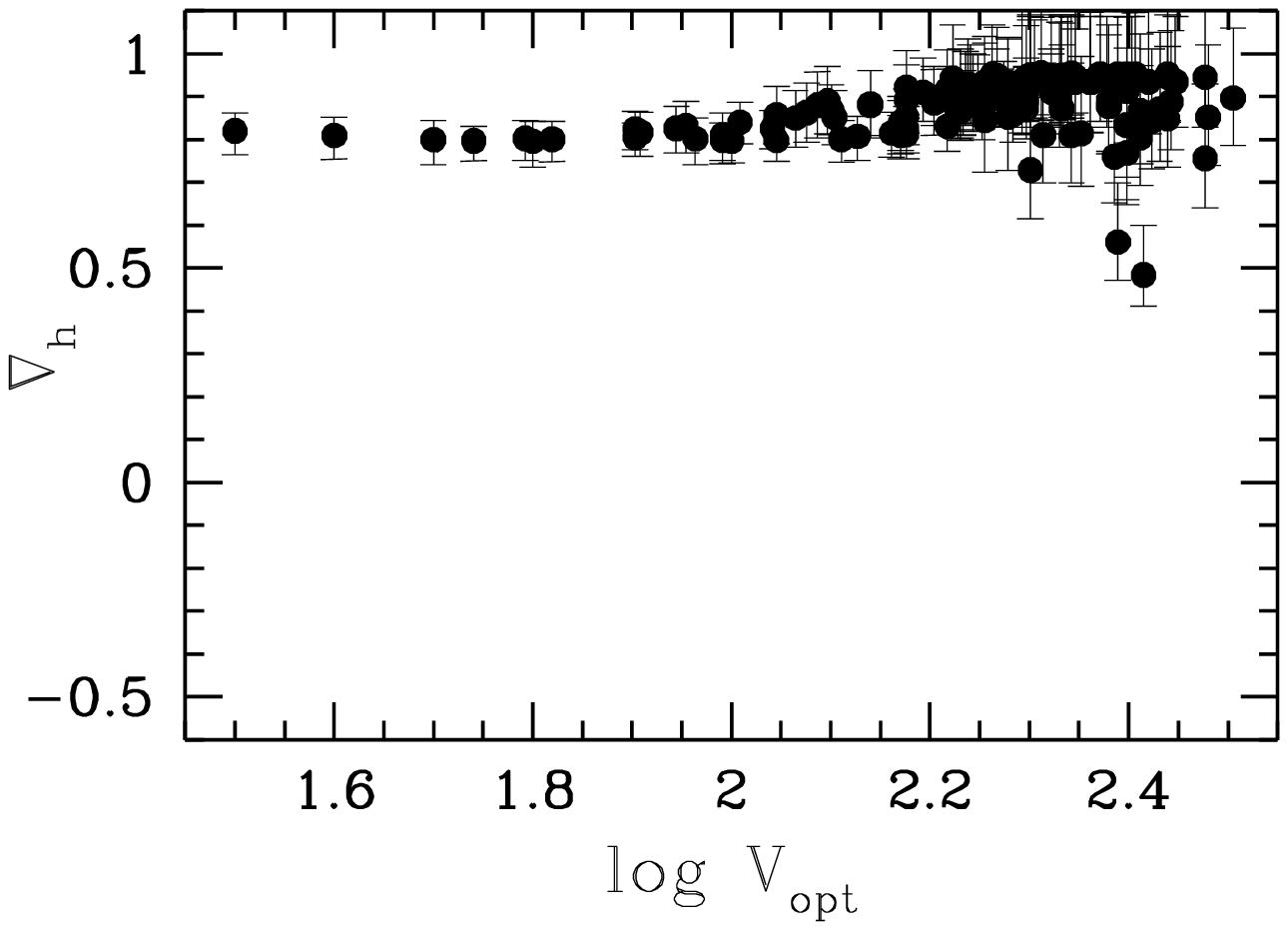}
\vskip -11.7cm
\hskip 8.2cm
\includegraphics[width=7cm]{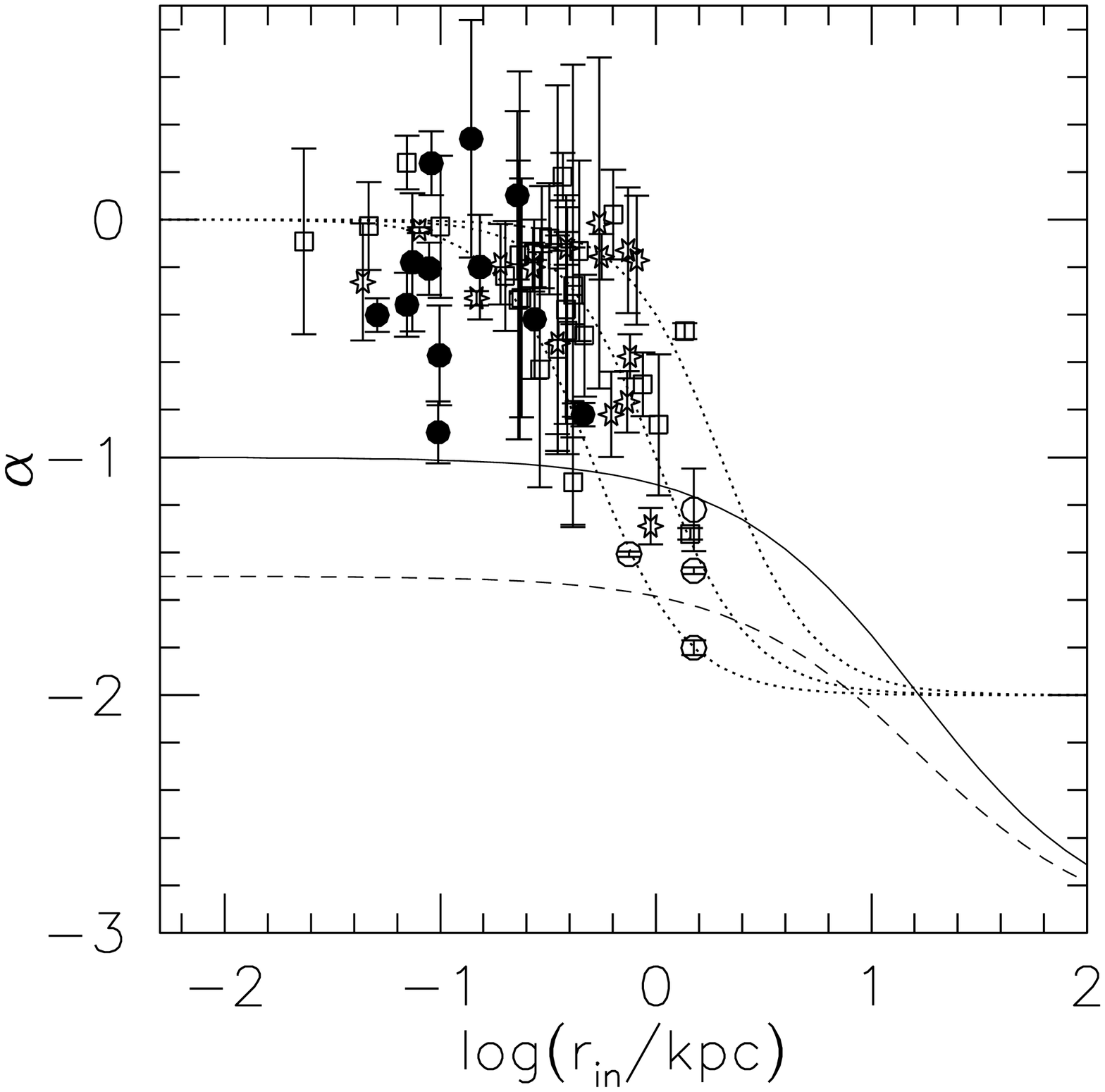}
\vskip -0.3cm
\caption{Left: DM  halo velocity slope $\nabla$ as a function of $V_{opt}$ \cite{Salucci01}, remind that:  $\nabla_{NFW} \leq 0.3$.
Right: Inner slopes of  LSB halo density  profiles {\it vs}  radii of the innermost data points \cite{deblok1}. Also  shown: pseudo-isothermal halo models  with core radii of 0.5, 1, 2  kpc (dotted lines) and  the NFW   profiles (full line)}
\vskip -0.2cm
\label{fig:nablaH}
\end{figure}

Presently,  there are  about ~100 spirals  whose RCs cannot be reproduced by a NFW halo + a stellar disk for any value of the model parameters. Furthermore, direct  investigation   has ruled out that an apparent core  may  arise from neglecting  certain kinematical effects \cite{gentile05}.
 
\begin{figure}[h!]
\centering
\vskip 1.cm
\includegraphics[width=6.5cm]{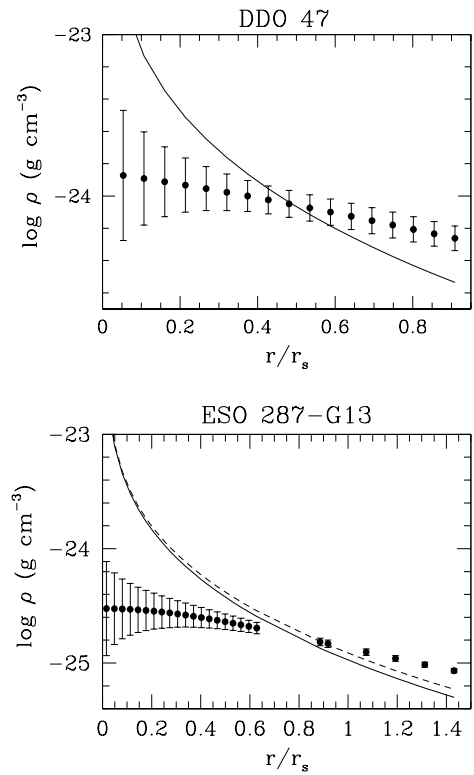}
\vskip -0.1cm
\caption{DM halo density of DDO 47 and ESO 287-G13. Solid lines: best fits for the NFW density profiles
that, for $0.7\leq r/r_s \leq1.3$ result {\it smaller } than the URC values (circles).
}
\label{fig:rho_47_287}
\end{figure}

A  complementary evidence   comes from  \cite{Salucci01} who derived,  in a model independent way, the logarithmic gradient of the {\it halo}  circular velocity $\nabla_h(r)\equiv \frac{d \log V_h(r)}{d \log r}$ at $R_{opt}$ in  140 spirals of different luminosity (see Fig. \ref{fig:nablaH}); their values  $\sim 1$  turned out to  independent of  galaxy magnitude  and inconsistent with  NFW halo predictions.  For  a large sample of  LSB a  similar result was obtained by  \cite{deblok1}, see Fig. \ref{fig:nablaH}).  
 
 Finally, accurate mass modeling of the external regions of 37 spirals with high quality RCs  led  to the discovery of further disagreement between data and  NFW  predictions \cite{Gentile07}.
 The DM halos around spirals have, in the inner regions,  densities  up to  one order of magnitude {\it lower}  than the $\Lambda$CDM predictions. At about $2.5 R_{opt} $ instead,  they  have densities   {\it higher}  by a factor 2-4  than the  corresponding NFW profile values (see Fig. \ref{fig:rho_47_287}).  DM  halos around spirals,  at ~5 kpc scale, are significantly  less dense than the  predicted  $\Lambda$CDM  halos, but  at ~50 kpc scale, they are denser than the latter. The DM  density core might be  associated to outward mass transfer.

\section{Conclusions and Remarks}

Considering a large sample of spiral galaxies we found that an  Universal Rotation Curve Model, that includes a cored  DM halo,   provides a very satisfactory fit to their Rotation Curves.  Cusped halo profiles,  a crucial feature of  standard $\Lambda$CDM  scenario,  are inconsistent with available kinematical data. Non standard features in such a  scenario might be able to reproduce the above discussed intriguing observations, however, this topic deserves a discussion that is  well beyond the goals of the present review.

The  success of the simple Universal Rotation Curve  model  \cite{Salucci07} in accounting for
the available kinematics   is  something notable. From a purely  empirical point of view, the distribution of luminous and dark matter  in galaxies shows amazing properties and a  remarkable systematics that are bound to  play a  decisive role in discovering  the nature of ``Dark Matter Phenomenon'' and in building a successful theory of  Galaxy Formation. 

A complete review on the topics dealt in this paper can be found at:  arxiv.org/abs/1102.1184

\end{document}